\documentclass[preprint,floats,aps,prl,floatfix,titlepage,tightenlines]{revtex4-2} 

\newcommand{\orcidicon}[1]{\href{https://orcid.org/#1}{\includegraphics[height=\fontcharht\font`\B]{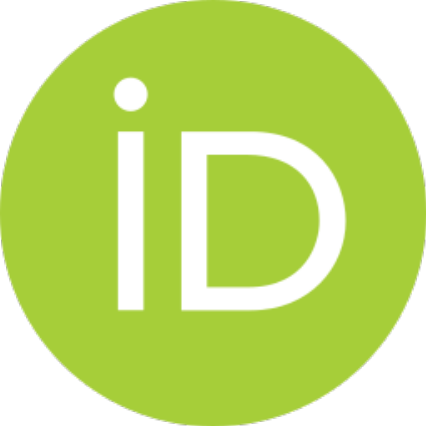}}}
\def\x{\mathbf{x}}

\def\g{\mathbf{g}}
\usepackage{graphicx}
\usepackage{graphics}
\usepackage{amssymb}
\usepackage{amsmath}
\usepackage{mathrsfs}
\usepackage[colorlinks=true,citecolor=blue,linkcolor=blue,urlcolor=blue]{hyperref}
\usepackage{tikz}
\usepackage{xcolor}
\usepackage{soul}
\usepackage{pstricks}
\usepackage{color}

\begin{document}

\title{Zero Point Density Fluctuations and Electron Brownian Motion}

\author{L. H. Ford\, \orcidicon{0000-0002-8359-120X}}
\email{Email address: ford@cosmos.phy.tufts.edu}
\affiliation{\mbox{Institute of Cosmology, Department of Physics and Astronomy, Tufts University,} \\ Medford, Massachusetts 02155, USA}

\author{G. O. Heymans\,\orcidicon{0000-0002-1650-4903} }
\email{Email address: olegario@cbpf.br}
\affiliation{Centro Brasileiro de Pesquisas F\'{\i}sicas - CBPF, \\ Rio de Janeiro, RJ 22290-180, Brazil}

\author{N.~F.~Svaiter\,\orcidicon{0000-0001-8830-6925}}
\email{Email address: nfuxsvai@cbpf.br}
\affiliation{Centro Brasileiro de Pesquisas F\'{\i}sicas - CBPF, \\ Rio de Janeiro, RJ 22290-180, Brazil}

\begin{abstract}
The  fluctuations in the phonon vacuum state can lead to zero point density fluctuations in a material, which in turn lead to local zero point
fluctuations of the dielectric properties of the material. We argue that the density fluctuations lead to a fluctuating force on a test charge, such
as an electron, located a short distance outside of the material. This force is due to fluctuating dipole moments inside the material, and produces
Brownian motion of the electron. We calculate the mean squared velocity of the electron in both the normal and transverse direction relative
to the boundary of the material. The result is nonzero in both directions, but larger in the normal case. We estimate the magnitude of this
quantum Brownian motion and find that, in some cases, it can exceed the effects  of both thermal motion and quantum  momentum
uncertainty.  This suggests that the effect may be observable, and could constitute a source of quantum noise in nanoscale devices.
 It also potentially offers a means to remotely sense zero point density fluctuations in a material.
\end{abstract}

\maketitle
 
The observability of zero point or vacuum fluctuations  is a subtle issue which has been discussed by several authors~\cite {Senitzky93,Gavish2000}.
Although it is generally agreed that the Lamb shift or the Casimir effect are observable consequences of vacuum fluctuations, it is often not clear
where the energy for a detection can come from if both the detector and the quantum system are in their ground states. However, both the Lamb shift 
and the Casimir effect can be viewed as energy shifts in a ground state energy which produce observable effects. The effect we will propose
can be viewed similarly.

One mechanism for detection of zero point fluctuations can be their effects on the Brownian motion of test particles. The dominant contribution to
the Lamb shift can be viewed as due {\blue to} the effect of electric field fluctuations on the motion {\blue of} an electron~\cite{Welton}. Several authors have investigated
Brownian motion in the modified vacuum fluctuations of a quantum field near a perfect mirror~\cite{YF04,Yu06,Seriu08,DeLorenci19,Guedes25} or in an expanding
universe~\cite{Bessa09}. Brownian motion has been invoked as a probe for the quantum nature of the gravitational field~\cite{F05}. Temperature
fluctuations in the cosmic microwave background due to gravitons from inflation have not yet been detected, but could  be viewed as arising from
photon Brownian motion in a bath of gravitons from inflation~\cite{Polnanev,KK16,FHW25}.

 Here we will deal with mass density fluctuations of a dielectric material due to zero point motion in the phonon vacuum state. Let
the density be a function of space and time of the form $\rho(\x,t) = \rho_0 +  \hat{\rho}(\x,t)$. Here  $\rho_0$ is the mean mass density,
and $  \hat{\rho}(\x,t)$ is an operator representing the quantum fluctuations around this mean value, which can be expanded in terms of
phonon creation and annihilation operators and plane wave  modes. This operator is analogous to that for a relativistic quantum scalar field,
but with the speed of light replaced by the speed of sound. This analogy has been exploited to study quantum field effects near black
holes~\cite{Unruh81,Unruh95} and in the early universe~\cite{FF04}.
The density fluctuations satisfy a correlation function~\cite{FS09}
    \begin{equation}\label{eq:dencor}
        \langle   \hat{\rho}(\x,t)   \hat{\rho}(\x',t') \rangle \hspace{-0.1cm}= 
        \frac{\hbar \rho_0}{2\pi^2 c_s} \frac{\Delta \x^2 + 3c_s^2 \Delta t^2}{( c_s^2 \Delta t^2-  \Delta \x^2)^3}\,,
    \end{equation}
where $c_s$ is the speed of sound in the dielectric and the expectation value is taken in the phonon vacuum state.
These density fluctuations are large enough to have been observed in light scattering experiments, as described in 
Refs.~\cite{FS09,Stephen69,VP76,Eramo,FS11}. See also Ref.~\cite{WF20}. 

\begin{figure}[htbp]
\includegraphics[scale=0.2]{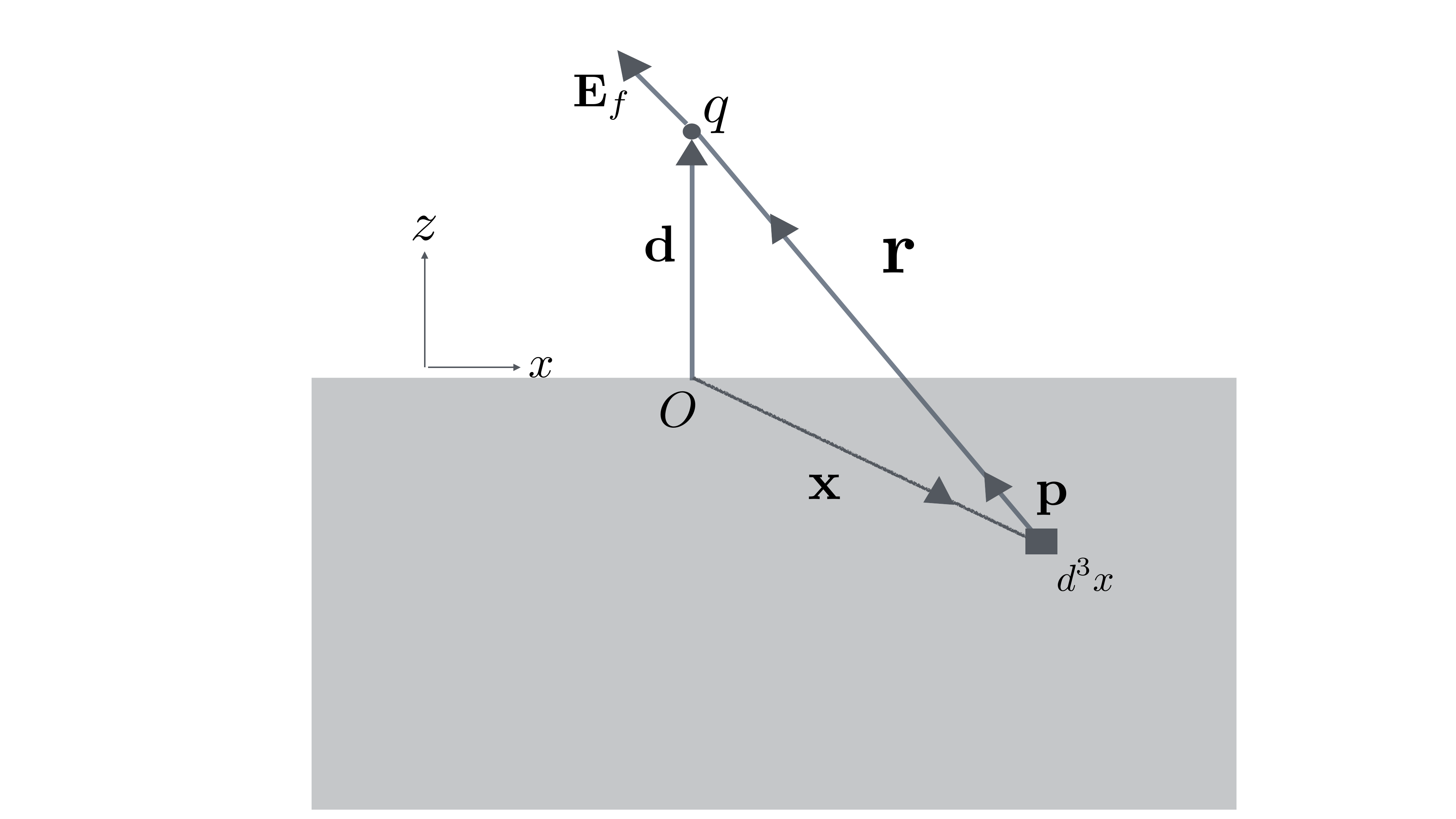}
\caption{A test charge $q$ is located a distance $d$ above a dielectric material with dielectric constant $\kappa$. Here the origin is at the point $O$.
The Coulomb field of the charge
induces a dipole moment ${\bf p}$ in a region with volume $d^3 x$ inside the dielectric. The electric field of this dipole exerts a force on the test
charge. The spatial integral of the time average of this force is the classical image charge force on the charge. Local density fluctuations produce a fluctuating part of the dipole moment and hence a fluctuating part ${\bf E}_f$ of the electric field acting on the charge. This fluctuating field in turn produces force and
velocity fluctuations of the charge, resulting in the Brownian motion. }
\label{fig:diagram}
\end{figure}

In this letter, we describe a different effect, in which the density fluctuations cause velocity fluctuations of a test charge, such as an electron, just
outside the material. The geometry is illustrated in Fig.~\ref{fig:diagram}.

The test charge $q$  located at $z = d$ outside the dielectric  induces polarization in the material. The total, time averaged, electric field of
this polarization exerts a force of the charge, which may be described as that due to an image charge $q' = - q (\kappa -1)/(\kappa + 1)$, 
located at $z = -d$~\cite{Jackson}.  Here  $\kappa$ is the dielectric constant of the material. The image charge force may equivalently be 
derived as the net force on the test charge due to induced polarization of the material. Here we assume that $\kappa$ is the static 
dielectric constant and ignore the effects of dispersion and dissipation in the material. At the frequencies of interest here, which are 
determined by the sound travel time over a distance $d$, this should be  a good approximation.

The effect of the zero point density fluctuations is to produce fluctuations of $\kappa$, and hence of the force on the test charge. Here we assume
small fractional fluctuations $ |\hat{\rho}| \ll \rho_0$. In this case, the dielectric constant has a small fluctuating part
\begin{equation}
 \kappa_1 = \frac{\hat{\rho}}{\rho_0} \, \kappa_0\,,
  \end{equation}
where $ \kappa_0$ is the mean dielectric constant, and $\kappa = \kappa_0 + \kappa_1$.

The electric field of $q$ inside the dielectric  is equivalent to that of a second image charge~\cite{Jackson} $q'' = 2\, \kappa \,q /(\kappa +1)$ at the location of the 
original charge, resulting in an electric field of ${\bf E}_q = - q\, {\bf \hat{r}}/[ 2\pi \epsilon_0 \, (\kappa + 1) \, r^2 ]$. This in turn produces a dipole
moment density of ${\bf P} = \epsilon_0\, \kappa\, {\bf E}_q $, and a dipole moment in volume $d^3 x$ of $d{\bf p} = {\bf P} \, d^3 x$. The electric
field of this dipole at the location of the test charge is
\begin{equation}
 d {\bf E}_d = \frac{3 \, {\bf \hat{r}}\, ( {\bf \hat{r}} \cdot d{\bf p}) -  d{\bf p} }{4 \pi \epsilon_0 \, r^3}
 = - \frac{q (\kappa -1) \, {\bf r} \, d^3 x}{4 \pi^2 \epsilon_0\, (\kappa + 1) \, r^6} \,.
  \end{equation}
If we set $\kappa = \kappa_0$ and integrate over the volume of the dielectric, the result is the electric field of the image charge, $q' $.

We are interested in the fluctuating electric field ${\bf E}_f$ due to the density fluctuations. If we Taylor expand to first order in $ \kappa_1$, the result
is
\begin{equation}
 \frac{\kappa -1}{\kappa + 1} =  \frac{\kappa_0 -1}{\kappa_0 + 1} +  \frac{2 \,\kappa_1}{(\kappa_0 + 1)^2}  + ... \,.
 \label{eq:Kappa-expand}
  \end{equation}
The contribution of region $d^3 x$ to ${\bf E}_f$ becomes
\begin{equation}
 d {\bf E}_f = - \frac{q \, \kappa_0  \, \hat{\rho} \, {\bf r} \, d^3 x}{2 \pi^2\, \epsilon_0\, \rho_0 \,(\kappa_0 + 1)^2 \, r^6} \,.
  \end{equation}
The net fluctuating field at the location of the test charge, after spatial integration,  is
\begin{equation}
 {\bf E}_f(t) =  - \frac{q \, \kappa_0} {2 \pi^2\, \epsilon_0\, \rho_0 \,(\kappa_0 + 1)^2}  \; \int   d^3 x  \, \hat{\rho}(\x,t)  \,   \g(\x)\, ,
  \end{equation}
 where
\begin{equation}
 g^j(\x) = \frac{r^j}{r^6} = \frac{d^j -x^j}{ | \mathbf{d} - \mathbf{x} |^6 } \,,
  \end{equation}
 with $\mathbf{r} = \mathbf{d} - \mathbf{x}$\,. Note that the electric field changes due to density fluctuations travel to the test charge at the speed
 of light, which is much larger than the speed of sound. As a result, we may view the response of the test charge to the density fluctuations 
 as instantaneous.

The electric field correlation function becomes
 \begin{equation}
 \langle E^j_f(t)\, E^k_f(t')\rangle =  \left[ \frac{q \, \kappa_0} {2 \pi^2\, \epsilon_0\, \rho_0 \, (\kappa_0 + 1)^2} \right]^2  \; \int   d^3 x \,  d^3 x'
 g^j(\x)\, g^k(\x')  \;  \langle   \hat{\rho}(\x,t)   \hat{\rho}(\x',t') \rangle \,,
  \end{equation}
  where $j$ and $k$ label Cartesian vector components.
The particle will undergo motion  in response to the fluctuating field. We will describe this motion using a Langevin equation, which takes 
the form of Newton's second law with a fluctuating force. Here we assume that an external force cancels the classical
image charge effect, so the particle responds to $ {\bf E}_f(t)$. This external force could come from an applied electric field, or as
a result of confinement in a structure, such as a nanotube.  If we set the initial velocity due to the zero point density fluctuations 
to zero, $v_{zp}^j(0) = 0$, then at time $t = t_1$,
$v_{zp}^j(t_1) = (q/m) \, \int_0^{t_1}  E_f^j(t) \, dt$, where $m$ is the mass of the particle. 
The mean squared speed in the $j$-direction at time $t_1$ is
\begin{equation}
 \langle (v_{zp}^j)^2 \rangle = \frac{q^2}{m^2}  \int_0^{t_1} dt \, \int_0^{t_1} dt' \; \langle E^j_f(t)\, E^j_f(t')\rangle\,. 
 \label{eq:v2-integral}
  \end{equation}
The field correlation function depends upon $\tau = t-t'$, so we may use the identity
\begin{equation}
 \int_0^{t_1} dt \, \int_0^{t_1} dt' \;  C(t-t') = 2 \int_0^{t_1} d\tau (t_1 - \tau)\, C(\tau) \,.
  \end{equation}
Here
\begin{equation}
  \langle   \hat{\rho}(\x,t)   \hat{\rho}(\x',t') \rangle \propto  C(\tau) = \frac{3\,  \tau^2 + \xi^2}{(\tau^2  - \xi^2)^3}
  \end{equation}
where $\xi = |\Delta \x|/c_s$. Now
\begin{equation}
 2 \int_0^{t_1} d\tau (t_1 - \tau)\, C(\tau) \sim \frac{1}{\xi^2}\,, \quad t_1 \gg \xi \, .
\end{equation}
Here we have evaluated the integration as an indefinite integral and then evaluated the result at the integration limits. Although $C(\tau)$ has a 
third-order pole at $\tau = \xi$, this procedure can be justified as either a Hadamard finite part, or as the result of analytically continuing $\xi$
into the complex plane before integration, and then back to the real line afterwards. 

This result tells us that the mean squared velocity components, $\langle (v_{zp}^j)^2 \rangle$, approach constant values at late times. The dominant 
contributions to the spatial integrations will come from  points separated by $|\Delta \x| \approx d$, so late times mean $t \agt d/c_s$. If the
electric field fluctuations were uncorrelated, we would expect to find the particle speed to undergo a random walk where 
$\langle v_{zp}^2 \rangle \propto t $, which would violate  energy conservation in the phonon vacuum state. The fact that this does not happen 
reflects strong anti-correlations in the density fluctuations and hence the electric field fluctuations. In effect, the electron can temporarily acquire
energy from a fluctuation, but an anti-correlated fluctuation will soon take back this energy. See Ref.~\cite{Parkinson11} for further discussion of 
this issue. The situation here is rather different from that in statistical physics, where dissipation is needed to constrain the growth of 
$\langle v^2 \rangle$ in Brownian motion.
Note that the density correlation function, Eq.~\eqref{eq:dencor}, is negative when $c_s \Delta t <  |\Delta \x|$, so
fluctuations outside the sound cone are anti-correlated. 

We can combine the above results to write an expression for the asymptotic values of  $\langle (v_{zp}^j)^2 \rangle$: 
\begin{equation}
 \langle (v_{zp}^j)^2 \rangle = A\, I^j \,,
  \end{equation}
where
\begin{equation}
 A = \frac{2 \hbar}{\pi^4\, c_s^3\, m^2\, \rho_0}\; \left(\frac{q^2}{4 \pi \epsilon_0}\right)^2  \;  \left[\frac{\kappa_0}{(\kappa_0 + 1)^2} \right] ^2 \,,
  \end{equation}
and
\begin{equation}
I^j =  \int d^3 x \, d^3 x' \; \frac{g^j(\x) \, g^j(\x')}{|\x -\x'|^2} \,.
  \end{equation}

The integral $I^j$ is proportional to $1/d^6$, so we can write $I^j = c_j/d^6$,where the $c_j$ are dimensionless constants, which may
be evaluated numerically, with the results~\cite{SM}
 \begin{equation}
 c_z \approx 0.68\,, \qquad c_x = c_y \approx 0.068\,.
  \end{equation}
Hence the mean squared speeds are
\begin{equation}
 \langle (v_{zp}^j)^2 \rangle = \frac{A\, c_j}{ d^6} \,.
 \label{eq:v2-result}
  \end{equation}
 We see that the  mean squared speed in the normal direction is about an order of magnitude larger than in a transverse direction. Note that
 we have assumed that the background mass density, $\rho_0$, is independent of position. The effects of spatial inhomogeneity, dispersion and
 dissipation in the dielectric will be topics for future work.
 
 We are now in a position to discuss several conceptual issues related to our model. The first is the question of the energy source for the Brownian motion.
 It may be viewed as arising from the classical electrostatic energy of the test charge in the image charge field ${\bf E}_d$. Recall that the fluctuating
 electric field  ${\bf E}_f$ which causes the Brownian motion is much smaller in magnitude. A  second question is whether the sudden switching at
 $t=0$ assumed in Eq.~\eqref{eq:v2-integral} alters the result, as was found in some other Brownian motion models~\cite{Seriu08,DeLorenci19}
 involving relativistic quantum fields. The
 following thought experiment suggests this is not the case here: Suppose that the switch-on occurred at a very large value of $d$ where both 
 $\langle (v_{zp}^j)^2 \rangle$
 and any effect of the switching are very small. Then suppose that the electron, or any structure containing it, are slowly lowered to a smaller value 
 of $d$. Consequently, the result given in Eq.~\eqref{eq:v2-result} is relatively independent of the details of the switch-on. This result may be viewed
 as a shift in the ground state energy of a system, just as in the cases of the Lamb shift and the Casimir effect. 
 
 There are two other sources of velocity fluctuations which are distinct from the Brownian motion effect treated here, but which need to be considered
 in an experiment. The first is the mean squared thermal velocity
at temperature $T$ in one direction, $v_T^2 = k_B\, T/m$, where $k_B$ is Boltzmann's constant. The other is the velocity variance,
$ \langle (v_{qu}^j)^2 \rangle$, due to the quantum momentum uncertainty of a localized quantum particle. If the electron is localized in
a region of length $\ell$ in direction $j$ , and the momentum uncertainty is of the order {\blue of the} uncertainty principle  minimum, then
\begin{equation}
 \langle (v_{qu}^j)^2 \rangle \approx  \frac{\hbar^2}{m^2 \, \ell^2}    \,,
  \end{equation}
 We may compare this effect with that of finite temperature using
 \begin{equation}
  \frac{\langle (v_{qu}^j)^2 \rangle}{v_T^2}  \approx  8.8 \,   \left(\frac{1\, K}{T}\right) \,  \left(\frac{10 {\rm nm}}{\ell}\right)^2  \,.
  \label{eq:q-estimate}
  \end{equation}
  
 It is useful to compare the numerical values of the zero point fluctuation induced speeds for an electron with the thermal and 
 momentum uncertainty effects using the above relation and
 \begin{equation}
 \frac{\langle (v_{zp}^j)^2 \rangle}{v_T^2} = \frac{9.1 \times 10^3\, c_j\, \kappa_0^2}{(\kappa_0 + 1)^4}\,  \left(\frac{1\, K}{T}\right) \, 
 \left(\frac{1 {\rm nm}}{d}\right)^6\,    \left(\frac{10^3 {\rm m/s}}{c_s}\right)^3\,  \left(\frac{10^3 {\rm kg/m^3}}{\rho_0}\right)\,. 
 \label{eq:estimate}
  \end{equation}
In the case of liquid $He^3$, where $T = 3.2 K$, $\rho_0 = 83 \, {\rm kg/m^3}$, $c_s = 100 \, {\rm m/s}$, and  $\kappa_0 = 1.026$, we have
  \begin{equation}
  \langle {(v_{zp}^j)^2 \rangle} \approx 8.6\,  c_j\,  \left(\frac{10 {\rm nm}}{d}\right)^6\, {v_T^2} \approx 
  3.1 \times 10^4\, c_j\,  \left(\frac{1 {\rm nm}}{d}\right)^6\,  \left(\frac{\ell}{ 1 {\rm nm}}\right)^2 \, \langle (v_{qu}^j)^2 \rangle  \,.
  \end{equation}
  In the case of liquid $N^2$, where $T = 77 K$, $\rho_0 = 808 \, {\rm kg/m^3}$, $c_s = 850 \, {\rm m/s}$, and  $\kappa_0 = 1.5$, 
  we find 
  \begin{equation}
 \langle {(v_{zp}^j)^2 \rangle} \approx 57 \, c_j\,  \left(\frac{1 {\rm nm}}{d}\right)^6\, {v_T^2} \approx 
      5.0 \, c_j\,  \left(\frac{1 {\rm nm}}{d}\right)^6\,  \left(\frac{\ell}{1 {\rm nm}}\right)^2 \,  \langle (v_{qu}^j)^2 \rangle  \,.
  \end{equation}
  We see that there are ranges of the parameters where the effects of zero point density fluctuations can dominate over both thermal motion
  and quantum momentum uncertainty. In the case of motion transverse to the dielectric interface, it is possible to have $\ell \gg d$, in which case 
  we can make $ \langle (v_{qu}^j)^2 \rangle$ very small. It is also possible to distinguish $ \langle (v_{zp}^j)^2 \rangle$ from 
  $ \langle (v_{qu}^j)^2 \rangle$ by their associated probability distributions. That for $ \langle (v_{zp}^j)^2 \rangle$ is Gaussian,
  as these are vacuum fluctuations arising from the ground state of the quantum harmonic oscillator. However, the distribution for
  $ \langle (v_{qu}^j)^2 \rangle$ depends upon the quantum state of the localized electron.

  Note the very strong dependence of $\langle (v_{zp}^2)^2 \rangle$ upon the distance of the electron from the boundary of the dielectric. 
  Our analysis, based upon Eq.~\eqref{eq:dencor}, strictly assumes a continuum description of the material, and hence $d$ large compared to the
  interatomic spacing. However, Eq.~\eqref{eq:estimate} may give a fair order of magnitude estimate even when $d$ is of the order of the
  interatomic spacing. 
  Another characteristic of our results is the marked anisotropy of the Brownian motion, in contrast to thermal motion. 
  
  Note that here we treat the fluctuations of the linear phonon operator $\hat{\rho}$, with the consequence that the electron velocity fluctuations
  will be described by a Gaussian probability distribution. This need not be the case if we included the effects of higher order terms in
  Eq.\eqref{eq:Kappa-expand}, which will be a topic for future work. A related topic is the possible correlation between light scattering by
  a localized density fluctuation~\cite{WF20}, and the electron velocity fluctuations. 
  
  If the electron is confined within a structure near the dielectric, such as a thin film or a nanotube, then our analysis can apply so long as the electron 
  Coulomb field and the induced dipole fields can penetrate the structure. In this case, the Brownian motion can act as a source of noise which may exceed
  the thermal noise and the effects of quantum momentum uncertainty.  Indeed, modern nanofabrication techniques routinely achieve electron confinement in ultrathin films, semiconductor heterostructures, and carbon nanotubes with characteristic dimensions of only a few nanometers, and in some cases approaching the $1\,{\rm nm}$ scale~\cite{Tans1998,Novoselov2004}. Since the velocity fluctuations induced by zero-point density fluctuations scale as $d^{-6}$, this regime is particularly favorable for their observation. As shown by Eqs.~\eqref{eq:estimate}, for distances of order $1\,{\rm nm}$ these fluctuations can become comparable to, or even exceed, both the thermal velocity fluctuations and those associated with quantum momentum uncertainty. Consequently, the effect discussed here may constitute an experimentally relevant source of quantum noise.
  
  Thus the effect treated here may be observable and have implications  both for nanotechnology  and for  fundamental physics, including as an analog model.
  
  \vspace{0.2in}
\begin{acknowledgments}  
G.O.H. thanks the Department of Physics and Astronomy at Tufts University for  hospitality while this work was performed. This research was supported by the Brazilian
agencies Fundação Carlos Chagas Filho de Amparo à Pesquisa do Estado do Rio de Janeiro - FAPERJ, CAPES (G. O. H.), Conselho Nacional de Desenvolvimento Científico e
Tecnológico - CNPq (Grant 305000/2023-3, N. F. S.),  and by NSF (Grant PHY-2207903, L.H.F.).
\end{acknowledgments}

\end{document}